\documentclass{nature}
\usepackage{amsmath}
\usepackage[utf8]{inputenc}
\usepackage{graphicx}
\usepackage{amsfonts}
\usepackage{bm}
\usepackage{hyperref}
\usepackage{bbold}
\usepackage{xcolor}
\usepackage[normalem]{ulem} %to strike the words

\usepackage[mathlines]{lineno}
\makeatletter
\let\saved@includegraphics\includegraphics
\AtBeginDocument{\let\includegraphics\saved@includegraphics}
\renewenvironment*{figure}{\@float{figure}}{\end@float}
\makeatother
\usepackage{ulem}

\title{Quantum Gravity Without Metric Quantization: From Hidden Variables to Hidden Spacetime Curvatures}

\author{Mohamed Hatifi}

\begin{document}
%\linenumbers
\maketitle

\begin{affiliations}
 \item Aix Marseille Univ, CNRS, Centrale Med, Institut Fresnel, UMR 7249, 13397 Marseille, France
 \item Quantum Machines Unit, Okinawa Institute of Science and Technology Graduate University, Okinawa 904-0495, Japan

\end{affiliations}
{Corresponding author: hatifi@fresnel.fr}

\begin{abstract}
    Bohmian mechanics offers a deterministic alternative to conventional quantum theory through well-defined particle trajectories. While successful in nonrelativistic contexts, its extension to curved spacetime—and hence quantum gravity—remains unresolved. Here, we develop a covariant extension of Bohmian mechanics in curved spacetime that removes the need for metric quantization. From a Lagrangian formulation, we derive a generalized guidance equation in which Bohmian trajectories generate “hidden curvature,” replacing metric superposition with a statistical ensemble constrained by Heisenberg uncertainty, offering a novel perspective on quantum gravity. Consequently, in our approach, measuring the gravitational potential at a point unveils a pre-existing trajectory and its associated curvature –a departure from the observer-centric paradigm of standard quantum mechanics--providing an alternative in which gravitational effects emerge from deterministic quantum trajectories rather than wavefunction collapse. Numerical simulations in Robertson–Walker and cigar soliton spacetimes reveal that while quantum interference is curvature-sensitive, Zitterbewegung remains invariant, distinguishing fundamental quantum effects. Moreover, deviations from the Born rule in inhomogeneous spacetimes are observed and suggest gravity-induced quantum non-equilibrium. This new approach has far-reaching implications for the role of determinism and potential observational signatures of quantum non-equilibrium in cosmology. 
\end{abstract}

\maketitle

\section{Introduction}

The quest to reconcile the principles of relativity with the framework of quantum mechanics represents one of the most exciting challenges in modern physics. Special relativity provides a natural framework for understanding the behavior of moving objects, while general relativity provides the spacetime medium in which objects move. It also describes, via Einstein's equations, how motions are affected by gravity in terms of curvatures of spacetime caused by mass and energy.  In contrast, quantum mechanics governs the behavior of particles at the smallest scales and stands, as well, as a fundamental pillar of modern physics. Although quantum mechanics governs phenomena at the smallest scales, its effects extend to macroscopic regimes, as evident in the cosmic microwave background \cite{huCosmic2002} and Hawking radiation from black holes \cite{hawkingRad1975}. However, extending quantum theory naively to larger scales leads to foundational challenges, such as the paradox of macroscopic Schrödinger cats, which lies at the heart of the measurement problem \cite{mermin1985,schlosshauer2005,martin2012,brukner2017}. In the standard interpretation of quantum mechanics, all physical information is encoded in the wavefunction, replacing definite trajectories with probabilistic measurement outcomes. This intrinsic indeterminacy is fundamentally at odds with the deterministic framework of general relativity unless either gravity is quantized \cite{mielnik1974,belenchia2018,smolin2000, MDM77a, Bose2017b,vedral, DowkerHensonSorkin} or quantum mechanics is "gravitized" \cite{penrose2014a, Penrose1996, diosi_gravitation_1984, diosi_models_1989, Jones1995, arxivSG}. In the past few years, many attempts have been made to quantize gravity through quantum field theory at different dimensionalities, and up to now, no one has succeeded in such a task \cite{deser1957,hawking1978,teitel1982,birrell1984,haag1984,hall2002, niedermaier2006}. In addition, the probabilistic formulation of quantum mechanics has sparked philosophical questions about the nature of reality and the role of consciousness and observation in the measurement process, which, if true, can have direct implications in cosmology \cite{mermin1985, aharonov1993, Durr95, Wfontology2002, esfeld2014b, ringbauer2015, CDW17,baccval}. Some physicists argue that quantum mechanics implies a fundamental indeterminacy in nature, while others contend that the wavefunction alone does not provide a complete description of a system\footnote{This perspective was notably expressed by Einstein, Podolsky, and Rosen in their seminal work \cite{EPR}, where they stated: \textit{While we have thus shown that the wave function does not provide a complete description of the physical reality, we left open the question of whether or not such a description exists. We believe, however, that such a theory is possible.}}. De Broglie and Bohm both advocated for this viewpoint and developed a deterministic interpretation of quantum mechanics \cite{db1924,dB1927,db1929,Bohm1952b,Bohm1952c} that restores the concept of a particle's position as a hidden variable that, jointly with the wave function, eventually explains the apparent randomness of quantum events and the dynamical emergence of the Born rule \cite{bohmvigier, Valentini91II, Valentini1992,Valentini2005a, Towler2011, Colin2012a, Abraham2014, Underwood2015, Colin2011, Hatifi2018b,hatifidrop, hatifithesis, hatifi2024_Springer}. In the nonrelativistic de Broglie-Bohm theory or the pilot wave interpretation of quantum mechanics \cite{db1924,dB1927,db1929,Bohm1952b,Bohm1952c}, the particles follow well-defined trajectories in time, which can be deduced from the guidance formula \({\bf \dot{x}}(t)=\frac{1}{m}\left.\nabla S\right|_{ x=x(t)}\), where \(S\) is proportional to the phase of the guiding wave \(\Psi\) and \(m\) is the particle's mass. 
% While the de Broglie-Bohm theory has proven effective in explaining non-relativistic phenomena, its intersection with the theory of relativity has posed considerable mathematical and conceptual challenges. Various approaches have been explored to extend Bohmian mechanics to relativistic frameworks, particularly in the context of special relativity. These efforts have involved modifications to the guidance equations and the incorporation of relativistic wave equations, such as the Dirac equation, to account for relativistic effects \cite{bohmTakabayasi1953,takabayashi1957,dbbQFT2004, nikolic2004,nikolic2005,nikolic2005b,colinDiracSea,braverman2013,durr2014,Durt2016, tumulka2018,nikolic2022,foo2022,fabbri2022,hatifi2024_PLA}.
Although the de Broglie-Bohm theory has successfully described non-relativistic quantum systems \cite{valentini-phd, Colin2010, sanz2014a, Abraham2014, Underwood2015, Hatifi2018b,hatifidrop, hatifithesis, hatifi2024_Springer}, its extension to relativistic settings remains a longstanding challenge. Reconciling Bohmian mechanics with relativity requires modifications to the guidance equations and a consistent treatment of probability currents in Lorentz-invariant wave equations. Several approaches have been proposed, particularly within special relativity, incorporating the Dirac equation and alternative formulations to accommodate relativistic effects \cite{bohmTakabayasi1953,takabayashi1957,dbbQFT2004, nikolic2004,nikolic2005,nikolic2005b,colinDiracSea,braverman2013,durr2014,Durt2016, tumulka2018,nikolic2022,foo2022,fabbri2022,hatifiPLA}.
\begin{figure}[h!]
\centering
\includegraphics[width=\textwidth]{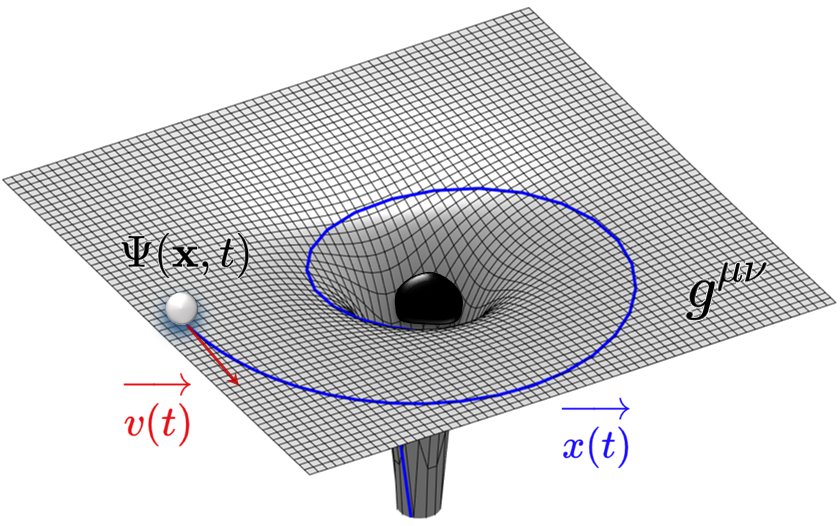}
\caption{
\textbf{Illustration of a Bohmian trajectory in curved spacetime near a black hole.} 
The grid represents the curvature of spacetime, described by the metric tensor \( g^{\mu\nu} \), highlighting the warping effect induced by the black hole's gravitational field. The blue line depicts the deterministic Bohmian trajectory, guided by the pilot wave \( \Psi({\bf x},t) \) through a generalized guidance equation \( u^\mu \). This schematic visualization emphasizes the interplay between quantum probability flow and the underlying spacetime geometry.
}
\label{fig:Bohmian_trajectory}
\end{figure}
However, the direct integration of Bohmian mechanics with curved spacetime and its explicit connection to general relativity has not been thoroughly addressed. The challenge lies in formulating a covariant theory that respects the geometric properties of curved spacetime while maintaining the deterministic trajectories central to Bohmian mechanics. Developing such a framework is crucial for several reasons. First, it could provide new insights into the path quantum particles take in strong gravitational fields, shedding light on the quantum aspects of black holes and cosmological phenomena. Second, it would contribute to the ongoing efforts to unify quantum mechanics and general relativity with new approaches, potentially revealing new fundamental physics principles. Lastly, a covariant Bohmian theory in a curved framework could offer an original point of view with practical applications in quantum gravity research and potentially help to bridge the gap between theoretical predictions and observational data.

To explore the implications of this framework, we first derive a generalized guidance equation from the Dirac Lagrangian in curved spacetime, establishing a covariant extension of Bohmian mechanics. We then apply this new formulation to two distinct geometries: the Robertson-Walker metric, representing an expanding and contracting universe, and the cigar soliton metric, which introduces localized curvature effects. These cases allow us to investigate how quantum probability flow adapts to different gravitational backgrounds and examine the role of curvature in modulating quantum interference and equilibrium. Building on these results, we propose a new trajectory-based extension of Einstein’s equations, where Bohmian trajectories act as local carriers of energy-momentum, dynamically influencing spacetime curvature. This approach offers an alternative perspective on the interplay between quantum mechanics and general relativity, providing a self-consistent framework that does not require metric quantization. Finally, we discuss the broader implications of this formulation, including its potential relevance to quantum gravity, black hole physics, and the emergence of cosmological structure.

\section*{Methods}

% \section*{Theoretical Framework}

\subsection{The Curved Spacetime:}
To establish the geometric framework in which the de Broglie-Bohm trajectories evolve, we begin by constructing the underlying spacetime background.  For analytical tractability and without loss of generality, let us consider first a (1+1)-dimensional curved spacetime and adopt natural units ($\hbar = c = 1$) with the metric signature \( \lVert \eta^{\mu\nu} \rVert = (+, -) \). In this convention, the mass \( m \) is dimensionless and effectively sets an inverse length scale. More generally, the fundamental idea developed here extends naturally to 3+1 dimensions. While higher-dimensional spacetimes introduce additional dynamical degrees of freedom, the core principles governing our approach remain unchanged. Moreover, despite its reduced dimensionality, (1+1)D gravity captures essential features of (3+1)D gravity while significantly simplifying mathematical complexity. As noted by Mann \cite{mann1991}, lower-dimensional models encapsulate key ideas of gravitational principles, making them theoretical tools allowing for possible deeper exploration of quantum effects in curved spacetime \cite{curvedDirac1994}. In (1+1) dimensions, the Einstein field equations reduce to a single scalar constraint equation rather than a dynamical system:
\begin{equation}
R - \Lambda = 8\pi G T,
\end{equation}
where $R$ is the Ricci scalar, $\Lambda$ is the cosmological constant, $G$ is the gravitational constant, and $T$ is the trace of the energy-momentum tensor $T=g^{\mu\nu}T_{\mu\nu}$. This equation is derived from a covariant action principle and ensures nontrivial curvature effects despite the reduced dimensionality. The low dimensionality makes it possible to derive analytical results explicitly while allowing numerical simulations to validate and illustrate important predictions, as we shall see.
\subsection{The Dirac Equation in Curved Spacetime}
To incorporate de Broglie-Bohm dynamics into curved spacetime, we propose to use the Dirac equation formulated in this setting \cite{carter1979,barut1987,curvedDirac1994,collas2019}. The Dirac equation in curved spacetime is expressed as:
\begin{equation}\label{diracurve}
\left( i \gamma^\mu({\bf x}) \nabla_\mu - m \right) \Psi ({\bf x},t) = 0,
\end{equation}
where $\gamma^\mu({\bf x})$ are the generalized gamma matrices in curved spacetime, $\nabla_\mu$ is the spinor covariant derivative, $m$ is the mass of the Dirac particle, and $\Psi({\bf x},t)$ is the 2-components spinor field. The metric tensor $g_{\mu\nu}({\bf x})$ relates to the Minkowski metric $\eta_{ab}$ through the \textbf{dyad} fields $e^a_\mu({\bf x})$:
\begin{equation}
g_{\mu\nu}({\bf x}) = e^a_\mu({\bf x}) e^b_\nu({\bf x}) \eta_{ab}.
\end{equation}
The dyads $e^a_\mu({\bf x})$ map vectors from the curved spacetime to the local inertial frame, facilitating the introduction of spinors in curved spacetime. Therefore, the generalized gamma matrices are defined by:
\begin{equation}
\gamma^\mu({\bf x}) = e^\mu_a({\bf x}) \gamma^a,
\end{equation}
where $\gamma^a$ are the constant Dirac gamma matrices in flat spacetime, satisfying the Clifford algebra:
\begin{equation}
\{ \gamma^a, \gamma^b \} = 2\eta^{ab}, \quad \sigma^{ab} = \frac{1}{2} [\gamma^a, \gamma^b], \text{ and}\quad \{ \gamma^\mu, \gamma^\nu\} = 2g^{\mu\nu}.
\end{equation}
The covariant derivative $\nabla_\mu$ acting on spinor fields includes the spin connection $\Gamma_\mu$:
\begin{equation}
\nabla_\mu = \partial_\mu + \Gamma_\mu,
\end{equation}
with the spin connection given by:
\begin{equation}
\Gamma_\mu = \frac{1}{2} \omega_\mu^{\,ab} \sigma_{ab}, \quad \text{where} \quad \omega_\mu^{\,ab} = e^{a\nu} \left( \partial_\mu e^b_\nu + \Gamma^\lambda_{\mu\nu} e^b_\lambda \right).
\end{equation}
Here, $\Gamma^\lambda_{\mu\nu}$ are the Christoffel symbols, defined as:
\begin{equation}
\Gamma^\lambda_{\mu\nu} = \frac{1}{2} g^{\lambda\rho} \left( \partial_\mu g_{\rho\nu} + \partial_\nu g_{\rho\mu} - \partial_\rho g_{\mu\nu} \right).
\end{equation}
Therefore, the Dirac equation \eqref{diracurve} in a $(1+1)$-dimensional curved spacetime simplifies to \cite{curvedDirac1994,collas2019}:
\begin{equation}\label{diracu}
\left[ i \gamma^a e^{\,\mu}_a \partial_\mu + \frac{i}{2} \gamma^a\,\frac{1}{\sqrt{-g}} \partial_\mu \left(\sqrt{-g} \,e^{\,\mu}_a\right) - m\,\mathbb{1}_2 \right] \Psi ({\bf x},t) = 0.
\end{equation}
where $\Psi({\bf x},t) = (\psi_1, \psi_2)^T$ is the two-component spinor field, and $g$ is the determinant of the metric tensor $g_{\mu\nu}$. In this reduced-dimensional framework, the spinor consists of two scalar components, corresponding to the spin-down (\( \psi_1 \)) and spin-up (\( \psi_2 \)) states of the particle. The flat Minkowski gamma matrices, \( \gamma^a \), can be conveniently expressed in terms of Pauli matrices. Throughout this paper, we adopt the conventions \( \gamma^0 = \sigma_1 \) and \( \gamma^1 = -i\sigma_2 \), where \( \sigma_i \) denotes the \( i \)-th Pauli matrix.  

After deriving the generalized guidance equation in curved spacetime, we shall illustrate the resulting trajectories for two distinct metrics, each representing a different type of universe:
\begin{align}
    \textbf{(1)} \quad ds^2 = dt^2 - a(t)^2\,dx^2, \quad \text{and} \quad 
    \textbf{(2)} \quad ds^2 = \tanh^2(x)\,dt^2 - dx^2.
\end{align}
The first metric corresponds to the spatially flat \textbf{Robertson-Walker metric} \cite{barut1987,stewart1990, curvedDirac1994,wands2000}, which describes an expanding or contracting universe governed by the scale factor \( a(t) \). The second metric is the \textbf{cigar soliton metric} \cite{witten1991,curvedDirac1994,lambert2012}, also known as the Witten black hole, which is particularly relevant for our study of quantum effects in near-horizon geometries. Although this metric emerges in lower-dimensional gravity, it shares key features with the Schwarzschild solution in (3+1)D, whose line element is given by:
\begin{equation}
    ds^2=\left(1-\frac{2M}{r}\right)dt^2-\left(1-\frac{2M}{r}\right)^{-1}dr^2-r^2\left(d\theta^2+\sin^2\theta d\varphi^2\right).
\end{equation}
The cigar soliton metric appears in effective lower-dimensional gravity models and has been studied in the context of two-dimensional dilaton gravity and conformal field theory. It is also related to black hole solutions in the SL$(2,\mathbb{R})/U(1)$ coset model, a two-dimensional conformal field theory relevant to non-critical string theory \cite{witten1991}. Additionally, it shares mathematical similarities with solutions in Jackiw-Teitelboim gravity \cite{Jackiw1984}, making it a useful model for exploring quantum gravitational effects in simplified settings.

% which describes ...and has been used in....
% \\
% \begin{equation*}
% \begin{aligned}
% \mathcal{L}_{\text{Dir}}\left(R_1,R_2,\varphi_+,\varphi_-\right)=&-\frac{1}{2} \sqrt{-g} \left[ 4 c^2 m R_1(t,x) R_2(t,x) \cos\left(\phi_m(t,x)\right) \right. \\
% &\quad + \hbar \left\{ R_1(t,x)^2 \left[ c (\epsilon_{10} + \epsilon_{11}) \left(\frac{\partial \phi_m(t,x)}{\partial x} + \frac{\partial \phi_p(t,x)}{\partial x}\right) \right. \right. \\
% &\quad \left. + (\epsilon_{00} + \epsilon_{01}) \left(\frac{\partial \phi_m(t,x)}{\partial t} + \frac{\partial \phi_p(t,x)}{\partial t}\right) \right] \\
% &\quad + R_2(t,x)^2 \left[ c (\epsilon_{10} - \epsilon_{11}) \left(\frac{\partial \phi_p(t,x)}{\partial x} - \frac{\partial \phi_m(t,x)}{\partial x}\right) \right. \\
% &\quad \left. \left. \left. - (\epsilon_{00} - \epsilon_{01}) \left(\frac{\partial \phi_m(t,x)}{\partial t} - \frac{\partial \phi_p(t,x)}{\partial t}\right) \right] \right\} \right]
% \end{aligned}
% \end{equation*}
\section{A generalized pilot-wave theory in curved space time}
% \section{A description {\bf{\it à la}} de Broglie-Bohm: a generalized guidance formula}
A relativistic quantum theory requires a well-defined equation of motion for trajectories that must be compatible with general covariance. To establish such a formulation, we start with a symmetrical Lagrangian density for the Dirac field~\eqref{diracurve} in terms of \( (\Psi, \bar{\Psi}) \):
\begin{equation}
    \mathcal{L}=\frac{i}{2}\sqrt{-g}\,\left[\overline{\Psi}\,\Gamma^{\mu}\,\nabla_{\mu}\Psi-\left(\nabla_{\mu}\overline{\Psi}\right)\,\Gamma^{\mu}\,\Psi+2i\,m\,\overline{\Psi}\Psi\right],
\label{lag}
\end{equation}
where \(\overline{\Psi}=\Psi^{\dagger}\gamma^{0}\) denotes the Dirac adjoint. A Lagrangian formulation ensures manifest Lorentz covariance, from which conserved quantities can be derived. From Noether’s theorem, this Lagrangian \eqref{lag} yields a conserved probability current, given by \(\nabla_\mu j^\mu = 0 \) with \(j^\mu=\overline{\Psi}\gamma^{\mu}\Psi\) which guarantees probability conservation throughout the curved spacetime. 
% Using Noether theorem, one can derive a conserved probability current \(\nabla_\mu j^\mu = 0 \) with \(j^\mu=\overline{\Psi}\gamma^{\mu}\Psi\) which explicitly gives the following components $j^{0} = |\psi_{1}|^{2}+|\psi_{2}|^{2}$ and $j^{1} = |\psi_{2}|^{2}-|\psi_{1}|^{2}$. Note that $(j^0)^2 - (j^1)^2 = 4 |\psi_{1}|^{2}  |\psi_{2}|^{2}\ge 0$ so that the current $j$ is necessarily timelike or null.

To obtain a guidance equation {\bf{\it à la}} de Broglie-Bohm that manifests explicitly through phase gradients, we rewrite each spinor component in terms of amplitudes and phases:
\begin{equation}
\Psi ({\bf x}, t) = \frac{1}{\sqrt{2}} e^{i S_+/2} \left(
\begin{array}{c}
R_1 \,e^{i S_-/2} \\
R_2 \,e^{-i S_-/2}
\end{array}
\right),
\label{eq:psi}
\end{equation}
where \( S_\pm = S_1 \pm S_2 \) and \( S_1 \) and \( S_2 \) are the quantum phases of the spinor components \( \psi_1 \) and \( \psi_2 \) respectively. Rewriting the Lagrangian density \eqref{lag} in terms of the current components $j^\mu$ and phase variables $S_+$, $S_-$  we obtain
\begin{equation}
\mathcal{L}=-m\sqrt{-g}\,(j_\mu j^\mu)^{1/2}\cos\left(S_{-}\right)-\frac{1}{2}\sqrt{-g}\,\left(j^{\mu}\nabla_{\mu}S_{+}+\epsilon^{\mu\nu}j_{\mu}\nabla_{\nu}S_{-}\right) 
    \label{lag2}
\end{equation}
where \( \epsilon^{\mu\nu} \) is the antisymmetric Levi-Civita tensor in two dimensions (\(\epsilon^{01} = -\epsilon^{10} = 1\)). Its presence ensures Lorentz invariance in \((1+1)\)-dimensional contexts and generates distinctive nontrivial coupling between the phase gradients and probability currents. Applying Noether’s theorem to the Lagrangian ${\mathcal L}(j^0, j^1, S_{+}, S_{-})$ in \eqref{lag2} for each variable gives the following additional set of dynamical and constraint equations governing the evolution of the system:
\begin{eqnarray}\label{setde}
\nabla_\mu j^\mu &=& 0\nonumber \\
\epsilon^{\mu}{}_{\alpha}\nabla_{\mu}j^{\alpha}&=&2m n \,\sin\left(S_{-}\right)\\
m\cos\left(S_{-}\right)\, j^{\mu}&=&-\frac{n}{2}\, (\nabla^{\mu}S_{+}+\epsilon^{\mu\nu}\nabla_{\nu}S_{-})\nonumber.
\end{eqnarray}
where \( n=\sqrt{j_\mu j^\mu}=2|\psi_{1}||\psi_{2}| \) represents the magnitude of the current $j^\mu$, ensuring that the current remains timelike or null in all cases. The set of equations \eqref{setde} generalizes previous findings in flat spacetime \cite{hatifiPLA}. This structure suggests a natural analogy with classical fluid dynamics, where \( j^\mu \)  plays the role of a particle current and \( n \) corresponds to the density of the quantum fluid. Following this perspective \cite{takabayasiRemarks1953, bohmTakabayasi1953,takabayashi1957, wallstrom,hatifiQW,hatifiPLA}, we define the velocity field \( u^\mu \) as the ratio of its current to its density \( u^\mu=j^\mu/n \). The evolution of the quantum trajectories in curved space-time is then governed by a generalized relativistic and covariant guidance equation, which, using \eqref{setde}, takes the form:
\begin{equation}
    m\cos\left(S_{-}\right)\, u^{\mu}=-\frac{1}{2}\, (\nabla^{\mu}S_{+}+\epsilon^{\mu\nu}\nabla_{\nu}S_{-}) \label{guidfc}
\end{equation}
This establishes a fully covariant generalization of the de Broglie-Bohm guidance equation consistent with relativistic quantum mechanics in curved spacetime. This formulation explicitly reveals the influence of curvature and quantum phase structure on particle trajectories. In the following section, we apply this formalism to specific gravitational backgrounds, investigating how Bohmian trajectories evolve in different spacetime geometries.

% \subsubsection*{Trajectory Computation}
% Initial conditions for particle positions and wavefunctions are chosen to match physically realistic scenarios. Particle trajectories are computed by integrating the guidance formula using a Runge-Kutta method. The Born rule statistics are analyzed by comparing the distribution of particle positions with the probability density given by \(|\Psi|^2\).

% \subsection*{Case Studies}

\subsubsection*{Robertson-Walker Spacetime}
\begin{figure}
    \centering
    \includegraphics[width=0.95\linewidth]{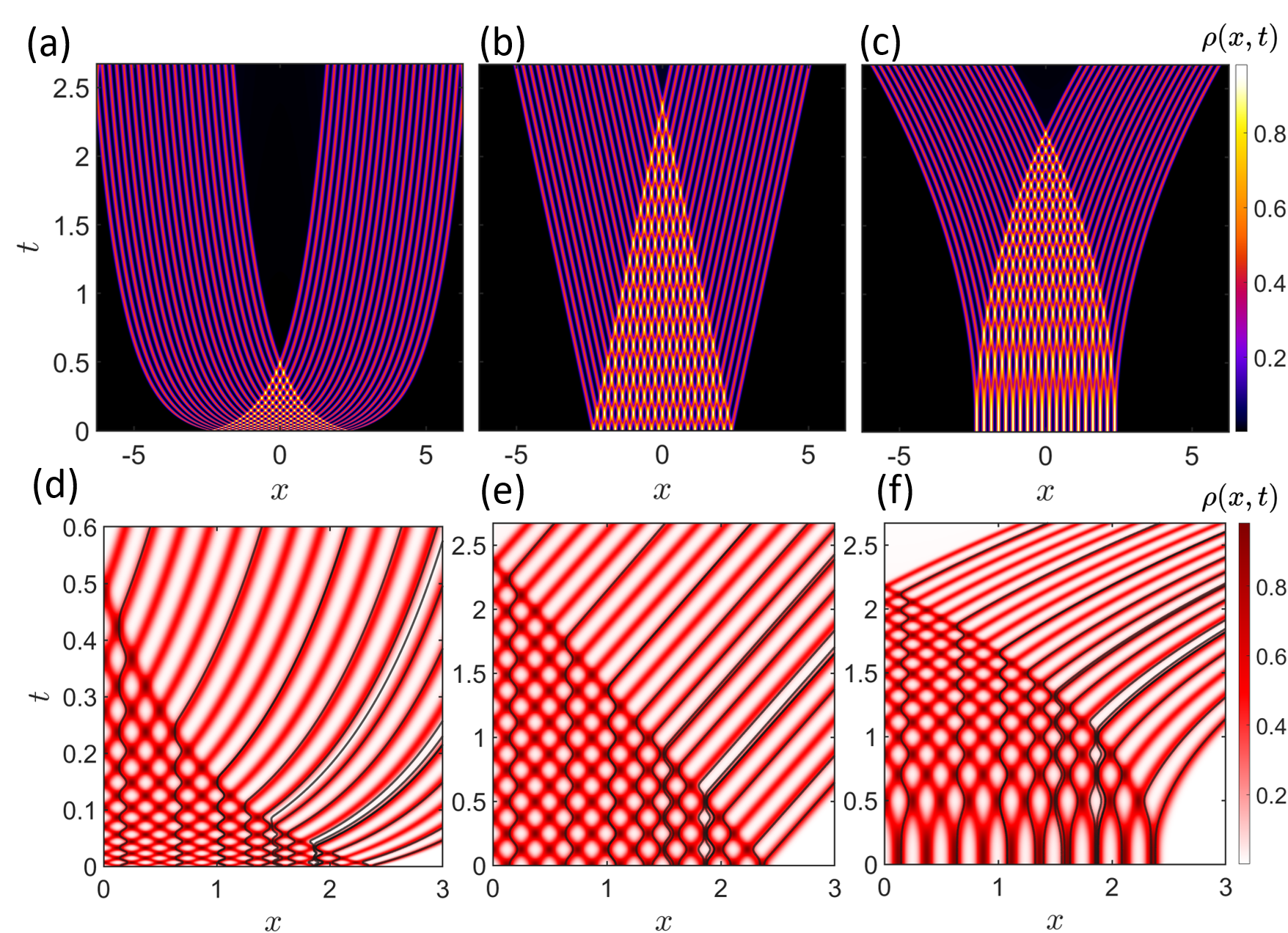}
%     \caption{
% \textbf{Quantum Probability Density and Bohmian Trajectories in Robertson-Walker Spacetime.}  
% \textbf{(a–c)} \textit{Quantum probability density}  $\rho(x,t)$ in spacetime for three different background geometries: \textbf{(a)} \textit{expanding universe} (\( k=1 \)), \textbf{(b)} \textit{flat Minkowski spacetime} (\( k=0 \)), and \textbf{(c)} \textit{contracting universe} (\( k=-1 \)). \textbf{(d–f)} \textit{Bohmian trajectories (solid lines) superimposed on the probability density}, illustrating the effect of curvature on quantum motion for identical initial conditions.  In the \textit{expanding universe} \textbf{(d)}, trajectories exhibit clustering due to cosmic expansion, whereas in the \textit{contracting universe} \textbf{(f)}, they disperse more rapidly. In the \textit{flat case} \textbf{(e)}, trajectories follow the expected quantum evolution in the absence of curvature effects. These results highlight the influence of spacetime expansion and contraction on quantum trajectories, leading to curvature-induced deviations from classical geodesic motion.}
   \caption{
\textbf{Bohmian Trajectories in Robertson–Walker Spacetime.} Panels \textbf{(a–c)} show numerical solutions for the relativistic quantum probability density, \( \rho(x,t) \), in three background geometries characterized by the scale factor \( a(t)=a_0t^k \): (a) an expanding universe (\( k=1 \)); (b) flat Minkowski spacetime (\( k=0 \)); and (c) a contracting universe (\( k=-1 \)). The deformation of \( \rho(x,t) \) captures the modulation of quantum interference by the evolving relativistic quantum potential. Panels \textbf{(d–f)} present Bohmian trajectories (solid lines) superimposed on the corresponding probability density, obtained by numerically solving the guidance equation \eqref{guidfc}. In all cases, trajectories are initialized at identical positions \( x_j(0) \) sampled from the initial probability distribution \( |\psi(x,0)|^2 \) in \eqref{psi0}. Curvature effects are evident: in the expanding universe (d), trajectories cluster as the metric stretches, whereas in the contracting universe (f), they disperse more rapidly; the flat case (e) serves as a comparison.
}

    \label{RWM1}
\end{figure}

To explore the implications of the dynamical equation \eqref{guidfc} in a relevant cosmological setting, we first apply it to the so-called Robertson-Walker metric \cite{barut1987,stewart1990, curvedDirac1994,wands2000}, which describes a spatially homogeneous and isotropic universe. This metric is widely used in modern cosmology to model expanding or contracting universes. It is described by the time-dependent scale factor function  $a(t)$ through $ ds^2 = dt^2 - a(t)^2\,dx^2$ and is characterized by a scalar curvature $R=g_{\mu\nu}R^{\mu\nu}=\ddot{a}/a$. From now on, we consider a power-law form for the scale factor, $a(t)=a_0\,t^k$. In this setting, the dyads are given by 
\begin{align}
    e^0_0=1,\quad e^1_1=\frac{1}{a(t)},\quad \text{and}\quad e^1_0=e^0_1=0
\end{align}
from which the Dirac equation \eqref{diracu} reads
\begin{equation}\label{d}
    i\,\left(\gamma^0 \partial_0+\frac{\dot{a}}{2a}\gamma^0+\frac{1}{a}\gamma^1 \partial_1 \right)\psi(x,t)-m\,\psi(x,t)=0
\end{equation}
In Figure \ref{RWM1}(a-c), we present the numerical solutions for the relativistic quantum probability density \( \rho(x,t)=\psi^\dagger\psi \) as a function of time for different cosmological backgrounds $a(t)=a_0\,t^k$: (a) an expanding universe $k=1$, (b) the flat case $k=0$, and (c) a contracting universe  $k=-1$, for comparison. These figures highlight the deformation induced by $a(t)$ on the probability density, stretching at the same time the relativistic quantum potential. To observe spacetime deformations under the time-dependent scale factor $a(t)$, we initialized the wavefunction with a state exhibiting interference patterns, and we looked at how the interferences are modified. Specifically, we considered an initial quantum state structured as a Talbot carpet, formed by a superposition of Gaussian wave packets:
\begin{equation}\label{psi0}
    \psi(x,0)=\frac{\mathcal{N}}{\sqrt{2}}\begin{pmatrix} 1 \\ 1 \end{pmatrix} \,\sum_j \exp \left({-\frac{(x-x_j)^2}{4\sigma_{x}^2}} \right)
\end{equation}
where $\sigma_x$ is the width of the gaussians , $x_j$ their initial positions and $\mathcal{N}$ is the normalization factor. In Figure \ref{RWM1}(d-f), we numerically solve the guidance equation \eqref{guidfc} and plot, on top of a color plot of $\rho(x,t)$, $14$ trajectories in full black lines for each spacetime scenario: (d) an expanding universe $k=1$, (e) the flat case $k=0$, and (f) a contracting universe  $k=-1$, for comparison. The initial positions $x_j(0)$ are identical across all cases (d-f) and were selected along the distribution $\vert\psi(x,0)\vert^2$.  By initializing $x_j(0)$ identically in each case, we isolate the effects of curvature on trajectory evolution. As expected for a formulation consistent with general relativity, the trajectories are inherently sensitive to the background geometry and dynamically respond to spacetime expansion and contraction.
\begin{figure}
    \centering
    \includegraphics[width=0.7\linewidth]{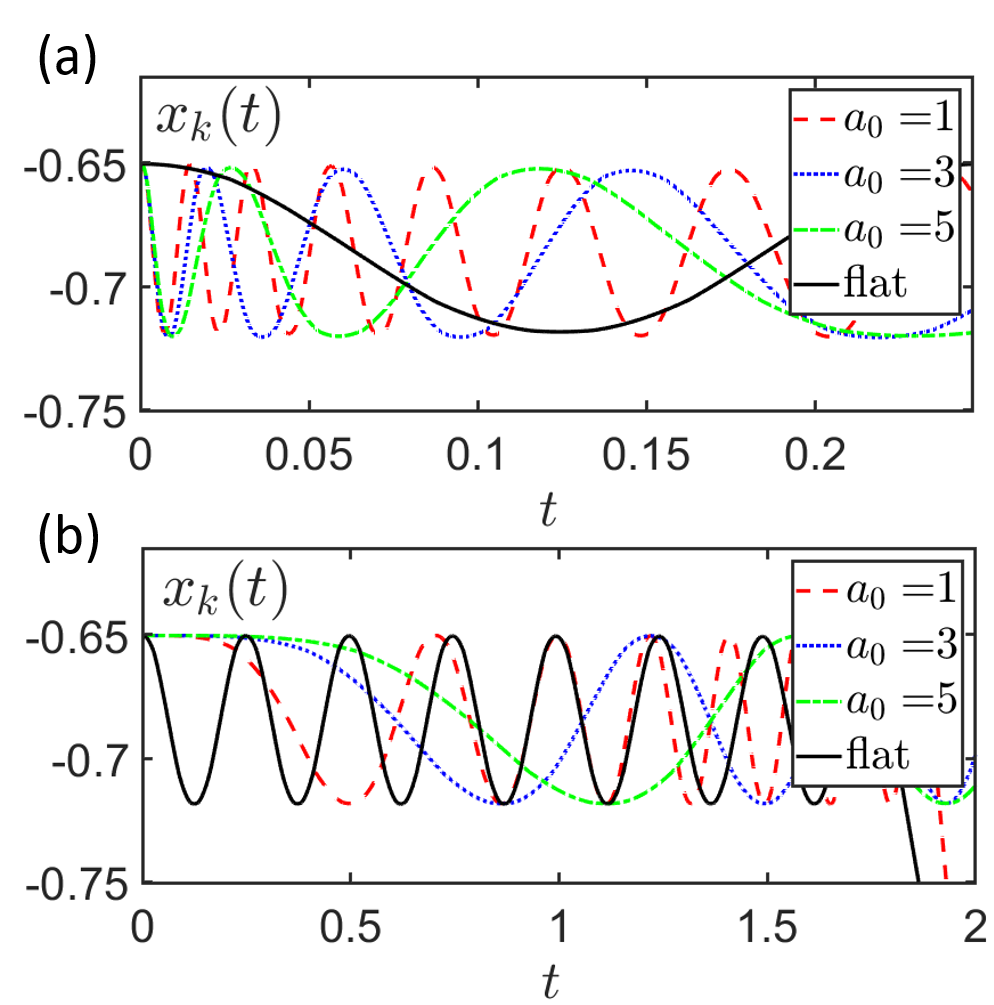}
    % \caption{In this figure, we plot a single relativistic de Broglie-Bohm trajectories \(x(t)\) for the Robertson-Walker metric associated with: (a) an expanding universe $k=1$ and (b) contracting universe $k=-1$. For each type of curvature we compare the trajectories obtained for different values of $a_0$ to the one corresponding to a pure flat space-time. For each trajectory, we started from the same initial condition whose distribution is described by the initial state defined in ..., and we considered a mass \(m=0.1\).}
    \caption{
\textbf{Influence of the Scale Factor on Bohmian Trajectories in Robertson-Walker Spacetime.} We show a single relativistic Bohmian trajectory \( x(t) \) in the Robertson-Walker metric with different values \( a_0 \) of the scale factor \( a(t)=a_0t^k \) for: (a) an expanding universe (\( k=1 \)) and (b) a contracting universe (\( k=-1 \)). The trajectory is initialized identically in all cases. A reference trajectory in flat spacetime is included for comparison. 
These results isolate the impact of spacetime expansion and contraction on a relativistic de Broglie-Bohm trajectory, demonstrating how variations in \( a_0 \) alter its evolution.
}

    \label{xtk}
\end{figure}

To better isolate the effects of spacetime curvature on individual trajectories, we plot in Figure \ref{xtk} the evolution of a single trajectory for $k=1$ and $k=-1$, considering different values of the initial scale factor $a_0$. The oscillatory behavior in the trajectories is not due to the Zitterbewegung but results from quantum interference provided by \eqref{psi0}, which is dynamically influenced by the evolving background geometry. In the expanding case (a), the increasing separation between quantum wavefronts leads to a gradual dephasing and spreading of the trajectory, reflecting the stretching of the underlying interference pattern. In contrast, in the contracting case (b), the interference fringes become more compressed, leading to a stronger trajectory localization, particularly for smaller values of $a_0$. 
\begin{figure}
    \centering
    \includegraphics[width=0.8\linewidth]{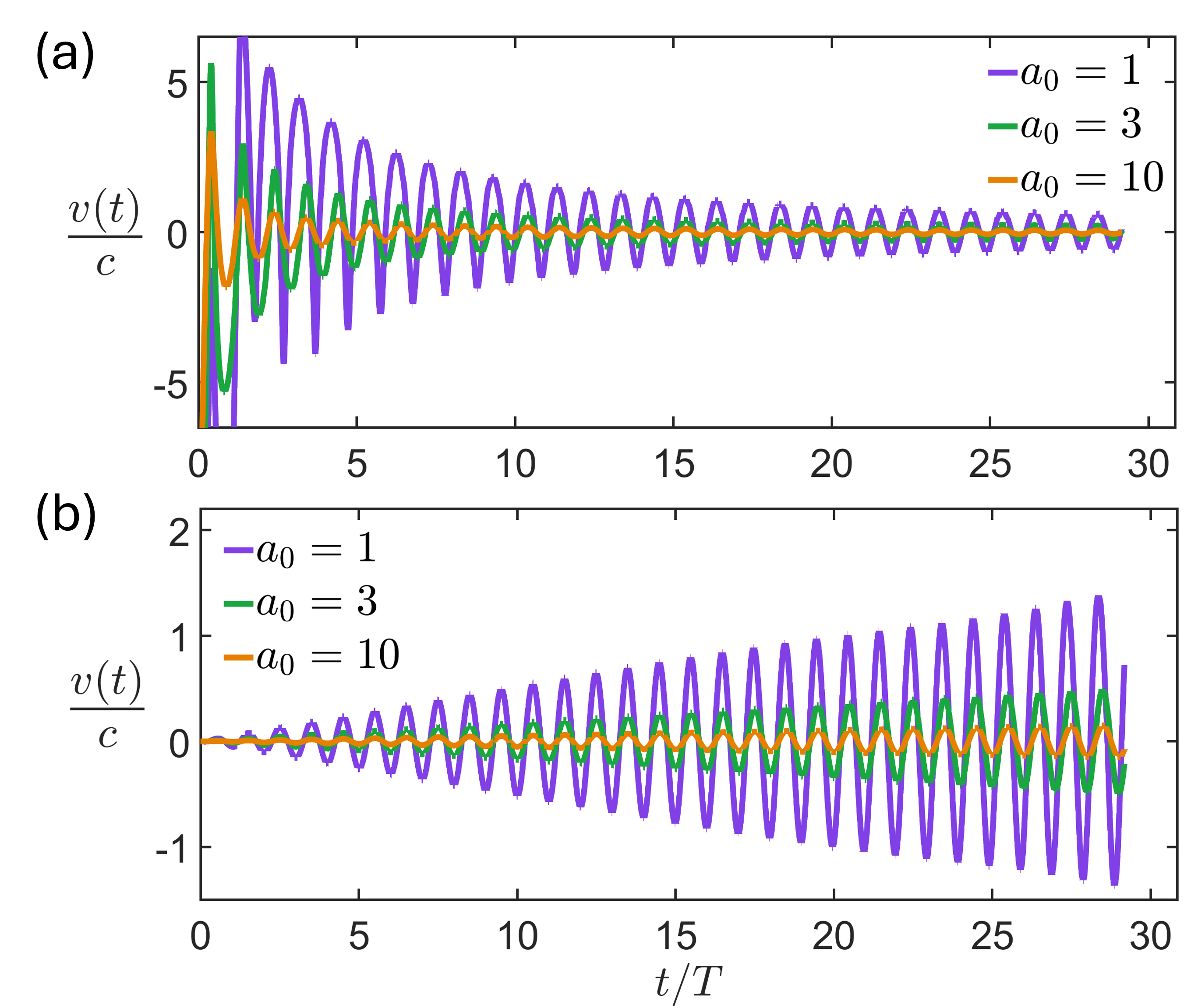}
    % \caption{\textbf{Zitterbewegung in curved spacetime.} Evolution of the Bohmian velocity \( v(t)/c \) for a single trajectory in an expanding universe $k=1$ (a) and a contracting universe $k=-1$ (b), for different scale factors \( a_0 \) and along the same trajectory. To isolate the Zitterbewegung effect, the particle mass was increased, given that the oscillation period scales as \( T = h / (m c^2) \). The results confirm that while the early-time behavior is influenced by metric evolution, the long-term Zitterbewegung oscillations remain unchanged across all backgrounds. Averaging the frequency over 30 periods verifies that Zitterbewegung is independent of \( a_0 \), reinforcing our model’s prediction that it is an intrinsic quantum effect unaffected by large-scale curvature. This highlights the robustness of our Bohmian framework in curved spacetime and suggests that fundamental quantum oscillations persist unaltered even in dynamically evolving universes.}
    \caption{
    \textbf{Zitterbewegung in curved spacetime.}  
    Evolution of the spatial component of the Bohmian velocity \( v(t)/c \) for a single trajectory in: (a) an expanding universe (\( k=1 \)) and (b) a contracting universe (\( k=-1 \)), for different values of the scale factor \( a_0 \). To isolate the Zitterbewegung effect, the particle mass was increased, as the oscillation period scales as \( T = h / (m c^2) \). Averaging the frequency over 30 periods confirms that Zitterbewegung is independent of \( a_0 \) in both cases and remains unchanged across all backgrounds, supporting our model’s prediction that it is an intrinsic quantum effect unaffected by large-scale curvature. This suggests that Zitterbewegung persists unaltered even in dynamically evolving universes. Despite early-time variations due to metric evolution, the long-term oscillatory behavior remains invariant.
    }

    \label{ZBa0}
\end{figure}
Another fundamental prediction of relativistic quantum mechanics is Zitterbewegung, the rapid oscillatory motion arising from the interference between positive- and negative-energy components of the Dirac spinor.  Figure \ref{ZBa0} presents the evolution of the spatial component of the Bohmian velocity \( v(t)/c \) for a single trajectory in both an expanding universe (a) and a contracting universe (b) for different values of \( a_0 \). In this case, the oscillations arise specifically due to Zitterbewegung rather than large-scale metric evolution. Unlike quantum interference, which is dynamically modulated by curvature, Zitterbewegung persists with the same characteristic frequency-- compton frequency defined by $T=h/(mc^2)$-- across all spacetimes, highlighting the consistency of our approach. A key outcome of our model is that it naturally predicts the invariance of Zitterbewegung under spacetime expansion and contraction, reinforcing its nature as a purely local quantum effect dictated by the intrinsic spinor structure.

It is also worth noting that the apparent superluminal behavior observed in the coordinate velocity of the Bohmian trajectories is not a violation of causality but rather a consequence of the time-dependent metric. For the Robertson-Walker metric, the relations $\gamma^t=\gamma^0$ and   $\gamma^x=\gamma^1/a(t)$ and thus $u^1\propto 1/a(t)$. This result explains the apparent divergence of the spatial velocity at early times in an expanding universe and at late times in a contracting universe, reflecting the influence of the evolving metric on the velocity flow. Precisely, the plotted velocity in figure \ref{ZBa0} corresponds to the spatial component and does not represent the speed measured by local inertial observers. Local measurements always involve the proper time and proper distance, which include factors of \( a(t) \), ensuring that the physical velocity remains bounded by \( c \).  
This is evident from the definition of the guidance equation: computing the norm of the probability current confirms that the scale factor cancels out, leaving an invariant quantity  

\begin{equation}
u^\mu u_\mu = g_{00}(u^0)^2 + g_{11}(u^1)^2=(u^0)^2 - a(t)^2(u^1)^2=c^2
\end{equation}

which remains independent of \( a(t) \). As a result, the physical (local) probability current—and therefore the associated velocity—remains causal, irrespective of cosmic expansion or contraction. The apparent \( v/c > 1 \) behavior in the coordinate representation arises purely due to the rescaling of spatial distances in the evolving metric and does not imply a breakdown of relativistic constraints when properly interpreted within the curved spacetime geometry.

Now, returning to Zitterbewegung, we isolate its effect by increasing the particle mass, as the Compton oscillation scales as $T=h/(mc^2)$. Our model predicts that, despite differences at short times due to metric evolution, the long-term oscillatory behavior remains invariant. By averaging the frequency over 30 periods in figure \ref{ZBa0}, we confirm that the Zitterbewegung frequency is independent in all cases, further supporting its insensitivity to large-scale curvature. The fact that this effect emerges naturally within our formulation strengthens the validity of our approach. Unlike conventional quantum field theory in curved spacetime, this provides a non-perturbative, trajectory-based insight into quantum evolution in curved backgrounds, offering a novel approach to understanding quantum effects in early-universe cosmology. These findings pave the way for further investigations into Bohmian quantum effects in extreme gravitational environments, such as near black holes. In the following section, we extend our analysis to the cigar soliton metric, which provides a relevant framework for exploring quantum trajectories in curved spacetime.

\section{Bohmian Quantum Dynamics in the Cigar Soliton Geometry}
\label{sec:CigarSoliton}

The cigar soliton metric,
\begin{equation}\label{cigmet}
    ds^2 \;=\; \tanh^2(x)\,dt^2 \;-\; dx^2,
\end{equation}
offers a compelling setting to explore the influence of a spatially localized curvature on quantum dynamics. This geometry, which arises naturally in certain lower-dimensional gravitational models as a limiting case of black hole spacetimes, is characterized by a pronounced horizon-like feature at $x = 0$ where the time component $g_{00} = \tanh^2(x)$ vanishes. The resulting extreme redshift in this region modifies particle motion in a manner analogous to the asymptotic behavior of infalling matter near a black hole horizon.
In contrast to the homogeneous cosmological backgrounds—such as the Robertson–Walker spacetime discussed earlier \cite{barut1987,stewart1990, curvedDirac1994,wands2000}, where the expansion or contraction of the universe reshapes the wavefunction globally—the cigar soliton geometry imposes a highly localized gravitational effect. This leads to an effective potential that confines wave packets near $x = 0$, as the modulation of time intervals by $\tanh(x)$ effectively slows down proper time evolution in the vicinity of the origin, analogous to how infalling matter asymptotically approaches a black hole horizon. As a result, Bohmian trajectories in this background are expected to exhibit constrained motion, with quantum probability currents dynamically adjusting to the varying curvature.

\begin{figure}
    \centering
    \includegraphics[width=0.9\linewidth]{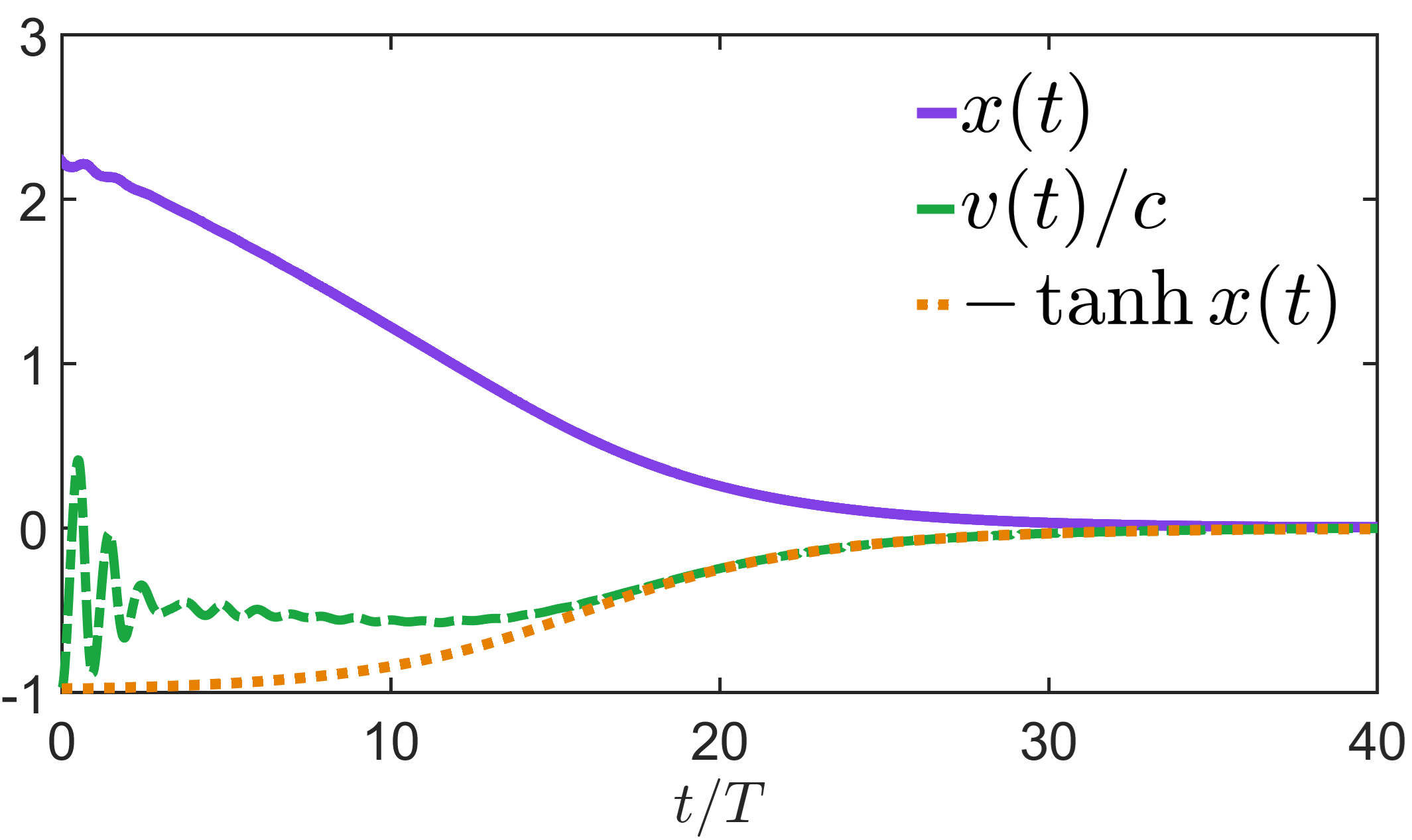}
    \caption{ 
    \textbf{Freezing of Bohmian trajectories near the horizon-like region in the cigar soliton metric.} The panels show the evolution of a single de Broglie-Bohm trajectory \( x(t) \), its velocity \( v(t) \), and the function \( \tanh(x(t)) \). The function \( \tanh(x(t)) \) is plotted as a reference since, for null geodesics in this metric, \( dx/dt = \tanh(x) \), providing a natural comparison between classical and quantum motion. As the trajectory approaches \( x \to 0 \), the velocity undergoes a rapid redshift, asymptotically vanishing, leading to a "freezing" effect near the horizon. This confirms how the metric dynamically constrains quantum motion in curved spacetime, highlighting a fundamental distinction between classical geodesics and Bohmian trajectories.}
    \label{fig:horizon_freezing}
\end{figure}

Adopting the dyad formalism as in the previous section, we define
\begin{equation}
    e^{0}_{\,0} \;=\; \coth(x), 
    \quad
    e^{1}_{\,1} \;=\; 1, 
    \quad
    e^{0}_{\,1} \;=\; e^{1}_{\,0} \;=\; 0,
\end{equation}
which leads to the Dirac equation in this curved background:
\begin{equation}\label{eq:DiracCigar}
    i \Bigl[\coth(x)\,\gamma^0 \,\partial_0 
    \;+\;\gamma^1\Bigl(\partial_1 
    \;+\;\tfrac{1}{\sinh(2x)}\Bigr)\Bigr]\psi(x,t)
    \;-\; m\,\psi(x,t)
    \;=\; 0.
\end{equation}
Here, the additional spin-connection term, $1/\sinh(2x)$, reflects the nontrivial curvature of the spacetime and ensures that the Dirac dynamics properly incorporate gravitational effects-- akin to a confining potential in non-relativistic quantum mechanics. 
\begin{figure}
    \centering
    \includegraphics[width=1\linewidth]{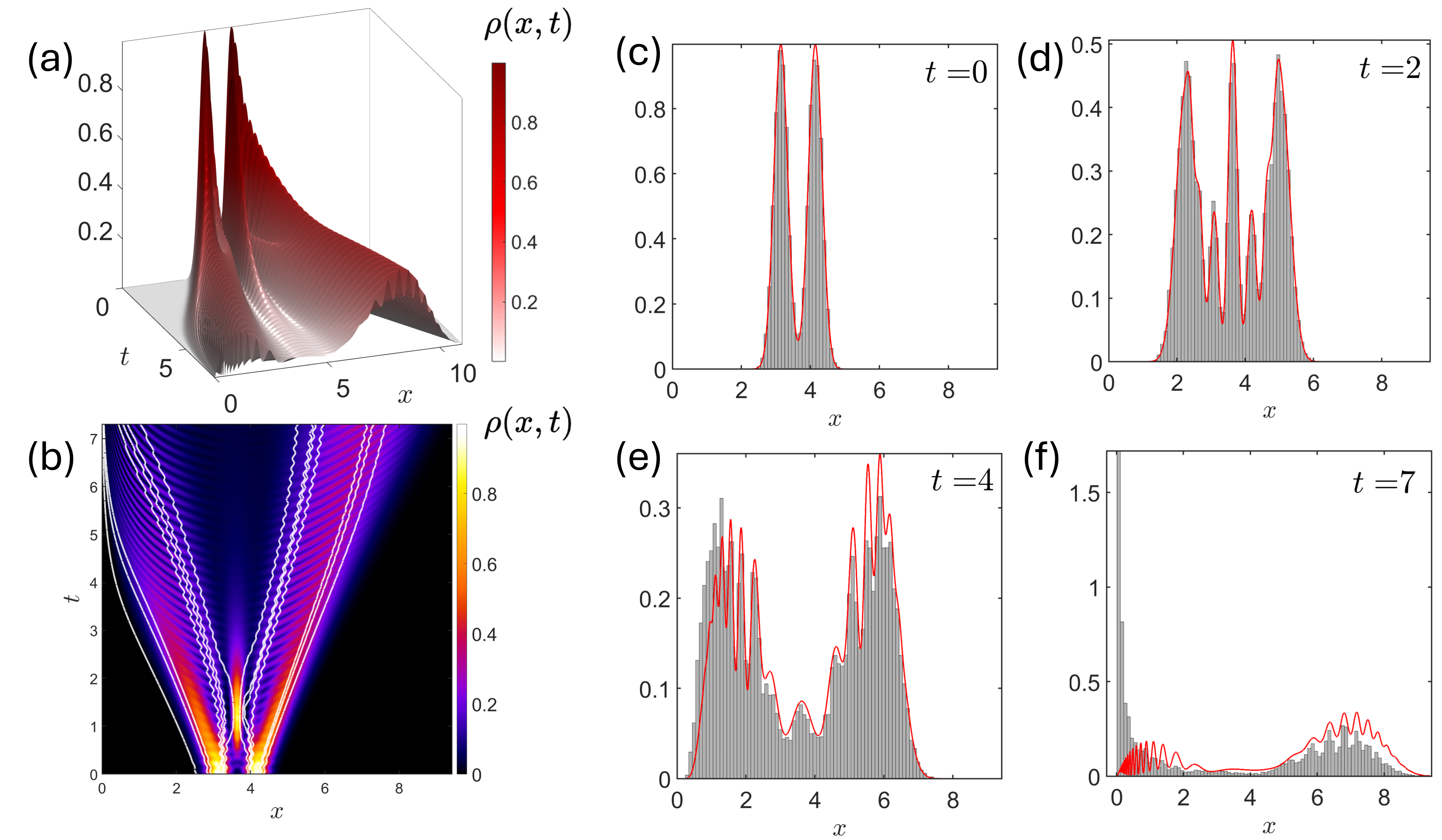}
    \caption{
\textbf{Bohmian Dynamics in a Cigar Soliton Spacetime.} (a) A three-dimensional plot of the quantum probability density, \(\rho(x,t)=\psi^\dagger\psi\), highlighting a pronounced modulation near \(x=0\). (b) Selected Bohmian trajectories (solid lines) are superimposed on a color map of \(\rho(x,t)\), obtained by integrating the relativistic guidance equation \eqref{guidfc} for the metric \eqref{cigmet}. Panels (c–f) display the temporal evolution of normalized histograms (bars) for the particle positions \(x_k(t_i)\) at representative times \(t_i\), overlaid with the corresponding probability density (red line). In these simulations, $20\,000$ trajectories were initiated from a superposed Gaussian \(\rho(x,0)\). The results reveal a dichotomy: trajectories for \(x>0\) exhibit dynamics close to free-particle behavior, whereas those approaching \(x=0\) experience a significant redshift—due to the vanishing \(g_{00}\) component—that effectively traps (or “freezes”) them near the horizon-like region.
}

    \label{fig:CigarSoliton}
\end{figure}
Figure \ref{fig:CigarSoliton} summarizes our numerical investigation. Panel (a) displays a three-dimensional representation of the quantum probability density, \( \rho(x,t)=\psi^\dagger\psi \), revealing a significant modulation near $x = 0$. In panel (b), selected Bohmian trajectories—obtained by integrating the relativistic guidance formula \eqref{guidfc} associated with the metric \eqref{cigmet}—are superimposed on a color map of $\rho(x,t)$. Additional panels (c–f) illustrate the temporal evolution of the normalized histograms of the distribution of positions $x_k(t_i)$ at a specific time $t_i$ alongside the quantum probability  \(\rho(x,t_i) \) in full red line. To make these histograms, we solved \eqref{guidfc} for $20\,000$ trajectories initially selected along the distribution $\rho(x,0)$ made of a superposition of two Gaussians. The numerical results reveal a dichotomous behavior of the quantum trajectories. Trajectories evolving in regions with large $x>0$ are only mildly affected by the curvature, exhibiting dynamics that closely approximate those of free particles in flat spacetime. Contrastingly, trajectories approaching $x = 0$ encounter a pronounced slowdown. This deceleration is directly attributable to the vanishing of the $g_{00}$ component at $x = 0$, which, as we said earlier, in turn, imposes an effective trapping potential. Consequently, the Bohmian velocities \eqref{guidfc} experience a significant redshift diminishing sharply as $x$ tends to 0, leading to an accumulation or “freezing” of trajectories near the horizon-like region. This behavior is illustrated in Figure \ref{fig:horizon_freezing}, where we track the evolution of a single Bohmian trajectory $x(t)$, its corresponding velocity $v(t)$, and the function $\tanh(x(t))$. The function $\tanh(x(t))$ is plotted as a reference since, for null geodesics in this metric, the equation $ds^2=0$ leads to $dx/dt=-tanh(x)$. This allows a direct comparison between classical geodesics and quantum trajectories. As the trajectory approaches the horizon-like region at $x\rightarrow 0$, the velocity undergoes a sharp redshift near the horizon, asymptotically diminishing toward zero. This phenomenon is analogous to the behavior observed in black hole spacetimes, where infalling matter appears to asymptotically hover near the horizon when viewed by a distant observer.
\\
This behavior is particularly striking when contrasted with flat spacetime, where the quantum equilibrium condition-- requiring that an initial ensemble distributed as $\mathcal{P}(x,0)=\rho(x,0) $ remains preserved at later times $\mathcal{P}(x,t)=\rho(x,t) $—is robustly preserved. In the cigar soliton geometry \cite{witten1991,curvedDirac1994,lambert2012}, however, we observe that while the quantum wavefunction avoids the singularity, the ensemble of Bohmian trajectories accumulates disproportionately near \(x=0\), leading to potential deviations from the Born-rule distribution \(\ \rho(x,t) \) in that region. This deviation emerges as a purely quantum-gravitational effect, absent in Minkowski spacetime, and induces $P(x,t)$ to develop fine-grained structures in that region that signal a localized breakdown of quantum equilibrium (Born rule). This finding underscores the profound sensitivity of quantum dynamics to localized curvature, revealing that spatial inhomogeneities in the metric can induce deviations from quantum equilibrium. Such effects suggest a potential mechanism for generating primordial asymmetries in the early universe, providing insight into the origin of cosmological background inhomogeneities \cite{huCosmic2002,Valentini2010b,pinto-neto_quantum_2013,colin2016b}.

Another important aspect concerns the role of nonlocality in Bohmian mechanics. While the wavefunction is globally defined and encodes nonlocal correlations, the evolution of individual Bohmian trajectories remains strictly local, being guided solely by the phase gradient of the wavefunction at each particle’s position. As a result, any curvature-induced modification—such as the pronounced redshift near $x=0$ in the cigar soliton geometry—only affects trajectories in the immediate vicinity. This ensures that while the wavefunction’s influence extends over all space, the clustering of trajectories near high-curvature regions arises purely from local modifications of the guidance equation without inducing instantaneous, nonlocal effects elsewhere. Consequently, the causal structure of the theory remains intact, precluding any superluminal signaling or acausal behavior.

This result opens intriguing directions for further investigation, particularly in the context of quantum entanglement in curved spacetime. A natural question is whether nonlocal correlations, which remain hidden in the single-particle trajectory dynamics, could manifest when considering entangled systems in strong gravitational fields. For instance, one could examine a scenario where one particle of an entangled pair falls into a black hole while the other escapes to asymptotic infinity. While standard quantum mechanics predicts the persistence of nonlocal correlations, it remains to be seen how Bohmian mechanics—where trajectories evolve deterministically—responds to such an extreme configuration. Investigating the interplay between nonlocality, entanglement, and curvature in this setting could provide valuable insights into the role of quantum mechanics in gravitational environments and its potential connection to semi-classical gravity.
\section*{Towards a Unified Framework}

The formalism developed in this work establishes a novel, trajectory‐based approach to quantum dynamics in curved spacetime via Bohmian mechanics. Although our analysis has been carried out in (1+1)-dimensional models for mathematical tractability, the core principles naturally extend to higher-dimensional settings. The use of simplified geometries has allowed us to extract key insights into the interplay between localized curvature and quantum probability flow.

In conventional treatments of quantum gravity, the gravitational field is often envisaged as a superposition of metric configurations, a notion that gives rise to deep conceptual challenges. In contrast, our approach posits that the metric remains unequivocally defined at all times, with its local curvature determined dynamically by the evolution of Bohmian trajectories. In this picture, the Einstein field equations
\begin{equation}
    G_{\mu\nu} + \Lambda g_{\mu\nu} = \frac{8\pi G}{c^4} \, T_{\mu\nu},
\end{equation}
acquire a new interpretation: the stress–energy tensor \(T_{\mu\nu}\) is not an externally imposed source but is constructed from the quantum matter dynamics as encoded by the Bohmian guidance equation. For example, for a spin–1/2 field, the conventional symmetric stress–energy tensor
\begin{equation}
    T^{\mu\nu} = \frac{i}{4} \left[ \overline{\Psi} \Gamma^\mu \nabla^\nu \Psi + \overline{\Psi} \Gamma^\nu \nabla^\mu \Psi - \left( \nabla^\mu \overline{\Psi} \right) \Gamma^\nu \Psi - \left( \nabla^\nu \overline{\Psi} \right) \Gamma^\mu \Psi \right]
\end{equation}
can be reexpressed within our framework as
\begin{equation}
    T^{\mu\nu} = \sqrt{-g} \left[ m\,n\,\cos(S_{-})\,u^\mu u^\nu + \frac{1}{2} \left( \epsilon^{\mu\alpha}\,j_\alpha\,\nabla^\nu S_{-} + \epsilon^{\nu\alpha}\,j_\alpha\,\nabla^\mu S_{-} \right) \right],
\end{equation}
where the first term represents the classical energy–momentum contribution of the Bohmian ensemble, and the second encapsulates quantum corrections emerging from the spinor phase dynamics. This reinterpretation directly couples quantum matter trajectories to spacetime curvature, offering an alternative route to quantum gravity that bypasses the need to quantize the gravitational field.

Importantly, while the global wavefunction is defined over all space and inherently nonlocal, the influence on the metric is mediated by the local distribution of Bohmian trajectories. Since the trajectory serves as a hidden variable, each quantum particle follows a well-defined path that carries with it a corresponding energy-momentum distribution, locally shaping the curvature of spacetime. However, due to Heisenberg uncertainty, the exact trajectory remains unknown, requiring an ensemble description where multiple possible curvatures coexist as a statistical representation of all admissible trajectories. Unlike standard approaches that assume a superposition of geometries, this framework suggests that measuring the gravitational potential at a given point in spacetime reveals a pre-existing Bohmian trajectory and, with it, the corresponding local curvature. In this view, spacetime is not in a quantum superposition but instead exhibits a well-defined but hidden curvature structure, emerging from the deterministic motion of quantum particles.
In other words, although the wavefunction carries nonlocal correlations, its effect on the curvature arises from the locally determined guidance equation--under what we studied in previous sections. This ensures that the dynamics of the metric respect causal propagation with no instantaneous nonlocal effects. In our formulation, any measurement of the gravitational potential at a given spacetime coordinate would reveal a pre-existing Bohmian position rather than generating new quantum states upon measurement. This represents a departure from the observer-centric paradigm of standard quantum mechanics, offering an alternative perspective in which gravitational effects emerge from deterministic quantum trajectories rather than wavefunction collapse.
Additionally, the proposed framework preserves the standard quantum uncertainty through the statistical distribution of trajectories, even as it provides a deterministic account of individual particle evolution. Moreover, by anchoring the gravitational field to the evolving Bohmian ensemble rather than to a superposed set of metrics, our approach may offer new insights into longstanding puzzles such as singularity resolution and the black hole information paradox.

Looking ahead, extending this formulation to incorporate additional quantum fields and to analyze entangled systems could yield further tests of its viability. In this manner, our work lays the foundation for a self-consistent, trajectory-based theory of quantum gravity that bridges the gap between classical spacetime and quantum matter.

\section*{Conclusion and Outlook}

In this work, we have developed a fully covariant extension of Bohmian mechanics to curved spacetime, yielding a deterministic, trajectory-based formulation of relativistic quantum dynamics. By applying this framework to models of expanding and contracting universes as well as to localized curvature in the cigar soliton geometry \cite{witten1991,curvedDirac1994,lambert2012}, we have demonstrated that Bohmian trajectories capture the imprint of spacetime evolution on quantum probability flow. Notably, while quantum interference is dynamically modulated by curvature, intrinsic effects such as Zitterbewegung remain unaltered, underscoring the robustness of local quantum features even in dynamically evolving backgrounds. Our formulation provides a novel route to quantum gravity without recourse to metric superposition. By expressing the stress–energy tensor in terms of the Bohmian guidance equation, quantum matter is seen to directly influence spacetime curvature via Einstein’s equations. In this picture, the gravitational potential at any point reflects the influence of pre-existing Bohmian trajectories rather than an indeterminate superposition of metrics. This deterministic approach offers fresh insights into the interplay between quantum mechanics and gravity and may help resolve foundational ambiguities associated with wavefunction collapse in gravitational settings.

Looking forward, several promising directions emerge. Extending the formalism to entangled quantum systems in strong gravitational fields—such as scenarios in which one particle of an entangled pair falls into a black hole while the other escapes—could yield new perspectives on quantum information transfer across event horizons and contribute to the black hole information debate. Moreover, the possibility of experimental tests in analog gravity platforms, ranging from trapped ions and superconducting circuits to graphene, opens a pathway to validate curvature-induced modifications of quantum probability currents. 
Finally, generalizing our approach to higher dimensions and incorporating non-Abelian gauge fields may lead to a more comprehensive trajectory-based framework for quantum field theory in curved spacetime. In particular, understanding the role of the quantum potential in semiclassical gravitational corrections and backreaction effects remains an intriguing open question. Whether this Bohmian approach can ultimately surmount the challenges of quantum gravity remains to be determined; however, our results establish a self-consistent, deterministic alternative that is fully compatible with relativistic covariance.

In summary, this work represents a significant step toward reconciling quantum mechanics with gravity. It opens a new perspective on how quantum matter and spacetime mutually interact, and it paves the way for further investigations into the foundational underpinnings of quantum gravity.

\section*{Data availability}
The data used for the current study are available from the corresponding authors upon reasonable request.

% \section*{Author Contributions}

\section*{Acknowledgements}
The author would like to thank R.I.N for invaluable support and encouragement. This work was supported by funding from the Okinawa Institute of Science and Technology Graduate University.

\section*{References}
% \bibliographystyle{unsrtnat} 
%\bibliography{main.bib}

\end{document}